\documentclass[usenatbib]{mn2e}
\usepackage{graphicx}
\usepackage{array}
\usepackage{multirow}
\usepackage[font=small,format=plain,labelfont=bf,up,textfont=normal,up]{caption}
\usepackage{float}

\title[Spectropolarimetric observations of 3C390.3]{Spectropolarimetric monitoring of active galaxy
3C390.3 with 6m telescope SAO RAS  in the period  2009-2014.
}

\author[Afanasiev et al.]{V. L. Afanasiev$^{1}$\thanks{E-mail: vafan@sao.ru},
 A.I. Shapovalova$^{1}$, L. \v C. Popovi\'c$^{2,3,4}$, N.V. Borisov$^{1}$\\
$^1$ Special Astrophysical Observatory of the Russian
Nizhnij Arkhyz, Karachaevo-Cherkesia 369167, Russia \\
$^2$ Astronomical Observatory, Volgina 7, 11060 Belgrade 74, Serbia \\
$^3$ Department of Astronomy, Faculty of Mathematics, University
of Belgrade, Studentski trg 16, 11000 Belgrade, Serbia\\
$^4$ Isaac Newton Institute of Chile, Yugoslavia Branch
 }

\begin{document}

\maketitle

\label{firstpage}

\begin{abstract}

Here we present the spectropolarimetric observations of the radio loud active galaxy  3C 390.3 in the period 2009-2014 
(24 epochs). The galaxy
has been observed with the
6-meter telescope of SAO RAS using  the SCORPIO spectropolarimeter.
We explore the variability and lags in the polarized light of the continuum and broad  H$\alpha$ line.
We give the Stokes parameters $Q, U$, degree of linear polarization $P$ and the position angle of the polarization
plane, $\varphi$, for 24 epochs.

We find a small lag~(10-40 days) between the unpolarized and polarized continuum
that is significantly  smaller   than the estimated lags for the unpolarized broad emission lines
(lag(H$\alpha$)$\sim$138-186 and lag(H$\beta$)$\sim$60-79 days). This  shows  that the {  region of the variable polarized continuum} is
significantly smaller than the broad line region, indicating that  a part of the polarized continuum
 is coming from the jet.
 The lag  of the polarized light in the H$\alpha$ line (89-156 days)
indicates an additional component to the disc one that has an outflowing velocity of $\sim$-1200 km s$^{-1}$.
This region seems to  depolarize the polarized broad H$\alpha$ line emitted from the disc and scattered in the inner part of the
torus.

\end{abstract}


\section{Introduction}

The radio loud active galactic nucleus (AGN) 3C390.3 (z=0.0561) belongs to a group of 10\% AGNs with  double peaked broad Balmer lines.
It is well known that 3C 390.3 shows a strong  variability (to $\sim$ 5-times)  in the UV/optical continuum and broad lines
\citep[see][]{ba80,ba83,yo81,nt82,pp84,cw87,vz91,zh96,wa97,di98,br98,sh01,sh10,se02,ta08,gu09,po11,di12}.
Also, 3C 390.3 has a high variability in the  X-ray spectrum, where a broad Fe K$\alpha$ line is present (Inda et al. 1994;
 Eracleous et al. 1996; Wozniak et al. 1998; Leighly et al.1997). The X-ray emission varies in a scale of several days, showing
 the highest variability in the lower energy region \citep[][]{gl03,gu09}, while the UV/optical broad lines vary at scales of
 several tens to hundred light days \citep[see e.g.][]{cw87,wa97,br98,di98,sh01,se02,sh10,di12}

Although the broad double-peaked emission lines of 3C390.3 are probably emitted from an accretion disc, there are some questions about the
nature of the broad line region (BLR), as e.g.
\cite{zh91} argued that a radial
biconical outflow is present in the BLR, but cross-correlation function (CCF) analysis shows that there is no
delay between the blue and red wings of H$\beta$, that
indicates  a dominant circular motion in the BLR \citep[][]{di98,sh01,sh10,di12,zh11,zh13}.
 However, besides two  peaks from the disc, there is an additional central, slightly red-shifted,  component in the
profiles of the broad  lines \citep[][]{sh10,po11}. Also, from time to time, there are
 some kind of perturbations in the disc \citep{jo10,po11}. Moreover, from a detailed study of the profiles of
the H$\alpha$ and H$\beta$ broad emission lines in the
monitored period (1995-2007), \cite{po11} have shown that the geometry of the BLR of 3C390.3 seems to be
very complex but the broad line region with
 the disc-like geometry has a dominant emission. On the contrary,
 spectropolarimetric observations \citep{co98,co00} indicate a BLR
 model in which the H$\alpha$ emission line is formed in the biconical flow and
 the polarized component of the line is scattered  on the inner part of the torus.
 The possible jet influence on the BLR and continuum emission has been discussed in \cite{ar07,ar10}. They found
 an observational evidence for the connection between the variable optical continuum of 3C390.3 nucleus,
  the compact radio emission of the jet on the sub-parsec scale and emissions of new observed jet components. To explain these correlations,
they suggested that the variable optical continuum emission originates
in the innermost part of the jet.

Note here that  the double peaked broad line profiles, in this and other double-peaked AGNs,
could be explained with  a number of
different  models for the BLR, as e.g. supermassive binary  black holes
\citep[][]{ga83,ga96,pop12}; outflowing biconical gas streams
\citep[][]{zh91};  in the accretion disc \citep[][]{pe88,ro92};
two-arm spiral waves in the accretion disc  \citep[][]{cw94};
or a relativistic eccentric (elliptic) disc \citep[][]{er95}, etc.

We performed the spectropolarimetric observations of 3C 390.3.
with 6m telescope of SAO in the period 2009-2014 period with the aim
to explore the disagreement between optical monitoring results which indicate dominant disc
like emission  \citep[see e.g.][and reference therein]{di98,sh01,sh10,jo10,po11,di12,zh11,zh13}
and spectropolarometric observations which are in favor of the biconical flow in the BLR
\citep{co98,co00}
Here we present the analysis of spectropolarimetric observations of 3C390.3 obtained in
24 epochs and  study how the parameters of the linear polarization
 changes with time in the broad H$\alpha$ emission line and continuum.

The paper is organized as following: in \S 2 we describe our observations and
data reduction, in \S 3 the results  are given, in \S 4 we discuss obtained results and
finally in \S 5 we outline conclusions.

\section{Observations and data reduction}

\subsection{Observation}

In the period 2009-2014   we performed spectropolarimetric monitoring of the broad line radio galaxy 3C 390.3
(24 epochs) with the 6-meter telescope of SAO RAS using the modified  spectrograph
SCORPIO \citep[][]{am05,am11} in the mode of the spectropolarimetry and  polarimetry in the spectral
range $4000-8000 \ \AA$ with the spectral resolution $4-10\ \AA$. We used two types of polarization
analyzers: the single Wollaston prism (WOLL-1) separates the ordinary and
extraordinary rays in two planes 0 and 90 degrees and a double
Wollaston prism (WOLL-2) consisting of two prisms illuminating half of
 the parallel beam and polarized light is separating planes in 0, 90, 45 and 135
degrees. For an unambiguous measurement of parameters of  the  linear polarization
in the first case it is required to obtain a sequence of four pairs of spectra at different angles of the phase plate
(0.45, 22.5 and 67.5 degrees), with WOLL-1 registering two long-slit spectra (with a slit 1-2$^{\prime\prime}$
width and 120$^{\prime\prime}$  height).
The second case records simultaneously four long-slit spectra  (60$^{\prime\prime}$  height) that uniquely
define the parameters of the linearly polarized radiation.
 For  calibration purposes, each night we observed the polarization of standard stars
 from the list of  \cite{hs82} and \cite{sm92}. Additionally, we observed non-polarized stars as spectrophotometric standards.
 We found that the accuracy of the linear polarization measurements is $\sim0.1\%$ \citep[more details in][]{,am05,aa12,af14}.
 The typical differences between our measurements of polarization standards and these in the catalogue
 were 0.1-0.2\% for polarization and
 2-4 degrees for the polarization angle.

In Table \ref{tab1} the log of observations is given, where are listed: Julian date, total exposure time,
number of polarization cycles, seeing, slit, analyzer,
grating, and spectral resolution. We observed in three different modes, using gratings VPHG550G, VPHG940, and VPHG1200 which
covered the H$\beta$ and H$\alpha$ wavelength ranges. Additionally we observed the inter-stellar matter
(ISM) polarization with the filter V at
 $\lambda$(max) 5500 \AA. The number of cycles denotes a number of observations for each position angle in the phase plate,
 i.e., one cycle corresponds to
the observations in all four above mentioned angles. The spectral resolutions given in Table
{  \ref{tab1}} were estimated using the Full Width at Half Maximum
(FWHM) of the lines from  the night-sky in the integral (AGN+night sky) spectra.

From Table \ref{tab1} it can be seen that there were large seeing variations between observations during the monitoring period.
{  This may cause changes in polarization
because the relative contribution of unpolarized host galaxy light will
change with the seeing,}
 however for this galaxy it is not a major problem, since the
emission of the nucleus is very bright compared to the host galaxy.

\begin{table*}
\begin{center}
\caption[]{Log of observations.}\label{tab1}
\begin{tabular}{ccccccccc}
\hline \hline
Date &JD&Total &Num. &Seeing&Slit&Analyzer&Grating/&Spectral \\
observation&240000+&exposure, s&of cycles&arcsec&arcsec&&filter&resolution (\AA)\\
\hline

2009.09.24 & 55099 & 3600 & 6 & 1 & 2.5 & WOLL-1 & VPHG550G & 20 \\
2010.07.17 & 55394 & 2400 & 5 & 2 & 2 & WOLL-1 & VPHG940 & 7 \\
2010.11.01 & 55501 & 3600 & 5 & 1.5 & 1 & WOLL-1 & VPHG940 & 5 \\
2011.05.01 & 55682 & 2160 & 3 & 2.5 & 2 & WOLL-1 & VPHG1200 & 6 \\
2011.06.01 & 55713 & 4000 & 5 & 1.6 & 1.5 & WOLL-1 & VPHG940 & 6 \\
2011.08.26 & 55799 & 1320 & 11 & 2.5 & 2 & WOLL-2 & VPHG940 & 6 \\
2011.09.27 & 55831 & 2400 & 5 & 1.6 & 2 & WOLL-1 & VPHG1200 & 5 \\
2011.11.20 & 55885 & 2880 & 6 & 2.5 & 1 & WOLL-1 & VPHG940 & 4 \\
2012.02.01 & 55959 & 1440 & 3 & 4 & 1 & WOLL-1 & VPHG940 & 4 \\
2012.02.14 & 55971 & 4320 & 6 & 3 & 2 & WOLL-1 & VPHG940 & 6 \\
2012.04.15 & 56032 & 2880 & 6 & 3 & 2 & WOLL-1 & VPHG940 & 6 \\
2012.05.17 & 56064 & 2400 & 5 & 1.2 & 2 & WOLL-1 & VPHG1200 & 5 \\
2012.06.21 & 56099 & 2880 & 6 & 1.3 & 2 & WOLL-1 & VPHG1200 & 5 \\
2012.06.21 & 56099 & ~600 & 15 & 1.5 & image & WOLL-1 & Jonson V& - \\
2012.08.24 & 56163 & 2160 & 6 & 1.2 & 2 & WOLL-1 & VPHG940 & 6 \\
2012.09.11 & 56181 & 2400 & 8 & 2.5 & 2 & WOLL-1 & VPHG940 & 5 \\
2012.10.07 & 56207 & 3600 & 8 & 2 & 2 & WOLL-1 & VPHG940 & 6 \\
2012.11.13 & 56244 & 2400 & 5 & 1 & 2 & WOLL-1 & VPHG940 & 7 \\
2013.02.06 & 56329 & 2400 & 8 & 3 & 2 & WOLL-1 & VPHG940 & 6 \\
2013.06.15 & 56458 & 4800 & 8 & 1 & 2 & WOLL-1 & VPHG940 & 6 \\
2013.11.03 & 56599 & 1260 & 21 & 3 & 2 & WOLL-2 & VPHG940 & 6 \\
2014.02.25 & 56713 & 3480 & 29 & 2 & 1 & WOLL-2 & VPHG940 & 4 \\
2014.03.06 & 56722 & 3600 & 20 & 2 & 2 & WOLL-2 & VPHG940 & 6 \\
2014.03.25 & 56741 & 3600 & 30 & 1.5 & 2 & WOLL-2 & VPHG940 & 5 \\
2014.05.30 & 56807 & 3600 & 30 & 2 & 1 & WOLL-2 & VPHG940 & 4 \\

\hline
\end{tabular}

\end{center}
\end{table*}
%
\subsubsection{Polarization of the ISM}

 The observed linear polarization of an object is a vector composition of the ISM polarization ($\vec{P}_{ISM}$)
 and polarization of the object ($\vec{P}_{AGN}$ -- in this case the radio-galaxy 3C390.3), 
 i.e. $\vec{P}_{obs}=\vec{P}_{AGN}+\vec{P}_{ISM}$.
The ISM polarization, as it is well known, depends on the Galactic latitude and it has  strong changes in the
rate of polarization and in the polarization angle  on one degree scale on celestial sphere. That is
connected with non homogeneous distribution of the ISM. In a number of papers,
the ISM polarization has been taken into account   as
a function of the Galactic extinction E(B-V)  for different latitudes as it is described in \cite{se75}.
However,
the problem with this method is that it does not take into account the direction of the ISM polarization vector,
 i.e., the vector
$\vec{P}_{ISM}$ has direction and intensity and both quantities should be taken into account \citep[see e.g.][]{kis04}.
Therefore, here we take into account the ISM polarization vector by measuring polarization of a number
of stars around the AGN, that  represents the ISM polarization. For these observations we
used the wide-field polarimetry,
as within the range of 0.45-0.8 mkm  a typical change in the ISM polarization
is smaller than 10\% from the maximum,
 i.e. for Galactic longitudes  $>25^\circ$ the estimate of  wide-field $\vec{P}_{ISM}$ is satisfactory
 for this purposes.

For the ISM polarization estimation we used  WOLL-1 analyzer with
rotated $\lambda/2$ phase plate, and with the 3-5$^\prime$  field. We
considered  only bright surrounding stars {within}  3 arcmin around 3C390.3  in the V filter.
We found 11 {brightest} stars in the field and for each star we calculated parameters
Q and U\footnote{For more detailed procedure of polarization parameters calculation in the image mode see \cite{ar14}.}.
We calculated for each of star  parameters Q and U and estimated the averaged polarization parameters of the
ISM to be: $Q_{ISM} = 0.64\pm0.23\%$ and $ U_{ISM} = -0.51\pm0.20\%$, $P_{ISM} = 0.82\pm0.22\%$ , $\varphi_{ISM} = 160.7\pm6.2^{\circ}$.
Further in the text we will use the  polarization parameters corrected
for the ISM polarization.

{ Note here that in the catalogue of \cite{he00} the rate of ISM polarization is
 changing around 0.5-1.1\% and angle of ISM
polarization around 159-163$^\circ$ in the radius of 5$^\circ$, that is not in contradiction with  our estimates.}

The contribution of the Inter Stellar Polarization (ISP) of the host galaxy of 3C390.3
has not been considered, since, as it is mentioned above,
the nucleus is significantly brighter than the host galaxy, and there is a problem to find the host galaxy ISP even using the
[OIII] lines, because in that case one assumes that these lines are
intrinsically unpolarized, which may not be the case.

\subsection{Data reduction}

The data reduction includes standard procedure for the long-slit spectroscopy, bias, flat field, geometrical correction along
the slit, correction of the spectral line curvature,  night-sky subtraction, spectral sensitivity of the instrument and
spectral wavelength calibration.
Additionally we used a comparison star in order  to remove  the strong atmospheric bands of O2 - B(6870\AA)
 and A(7600\AA) from 3C390.3 spectrum. Note that the B band is on the blue wing of H$\alpha$. We
 integrated spectra along the slit, since the procedure
 of decomposition (of the observed light as a function of wavelength along the slit) of the host galaxy and AGN
  increases the statistical errors. integration interval was $\pm (10^{\prime\prime}-15^{\prime\prime})$. This systematic error due to
wings image does not exceed 0.1\% in the value of the degree of polarization.

Values of normalized Stokes  parameters  $Q(\lambda)$, $U(\lambda)$ and total intensity $I(\lambda)$  for
WOLL-1 analyzer can be found from the simple relations \citep[see also][]{af14}:
$$Q(\lambda)={1\over{2}}\biggl({I_{0}(\lambda)-I_{90}(\lambda)\over{I_{0}(\lambda)+I_{90}(\lambda)}}\biggr)_{\phi=0}-
{1\over{2}}\biggl({I_{0}(\lambda)-I_{90}(\lambda)\over{I_{0}(\lambda)+I_{90}(\lambda)}}\biggr)_{\phi=22.5},$$

$$U(\lambda)={1\over{2}}\biggl({I_{0}(\lambda)-I_{90}(\lambda)\over{I_{0}(\lambda)+I_{90}(\lambda)}}\biggr)_{\phi=0}-
{1\over{2}}\biggl({I_{0}(\lambda)-I_{90}(\lambda)\over{I_{0}(\lambda)+I_{90}(\lambda)}}\biggr)_{\phi=67.5},$$

$$I(\lambda)=\sum_{\phi}[I_0(\lambda)+I_{90}(\lambda)]_{\phi},~~~~~~~~~\phi=0,45,22.5,67.5$$
For WOLL-2 analyzer appropriate values are determined from the relations:
$$Q(\lambda)={I_{0}(\lambda)-I_{90}(\lambda)\over{I_{0}(\lambda)+I_{90}(\lambda)}},$$

$$U(\lambda)={I_{45}(\lambda)-I_{135}(\lambda)\over{I_{45}(\lambda)+I_{135}(\lambda)}},$$

$$I(\lambda)=I_0(\lambda)+I_{90}(\lambda)+I_{45}(\lambda)+I_{135}(\lambda)$$

Then we calculated the degree of the linear polarization $P(\lambda)$ and angle of polarization plane $\varphi(\lambda)$ as:
$$P(\lambda)=\sqrt{Q(\lambda)^2+U(\lambda)^2} \ \ \ \ \varphi(\lambda)={1\over 2}{\rm arctg}[U(\lambda)/Q(\lambda)]$$

Using the procedure of the polarization parameter calculation from \cite{aa12}
we  estimated  the Stokes parameters ($I(\lambda)$, $Q(\lambda)$ and $U(\lambda)$),
 the degree of the linear polarization $P(\lambda)$ and position angle of
polarization plane $\varphi(\lambda)$ for the rest wavelength 5500 \AA.
Also we determined  the continuum flux  and the shift of the broad emission
line in the polarized light  relatively to the systematic velocity  $\Delta V=V_{pol}-V_{sys}$.
In further analysis, the  polarization parameters ($I, Q, U,P~ \rm and ~\varphi$) have been robustly estimated
as average in a spectral window of 25-30 \AA ~for all cycles of measurements,
the number of the windows for different observation data are between 5 and 10.

\begin{figure}
\centering
\includegraphics[width=8.3cm]{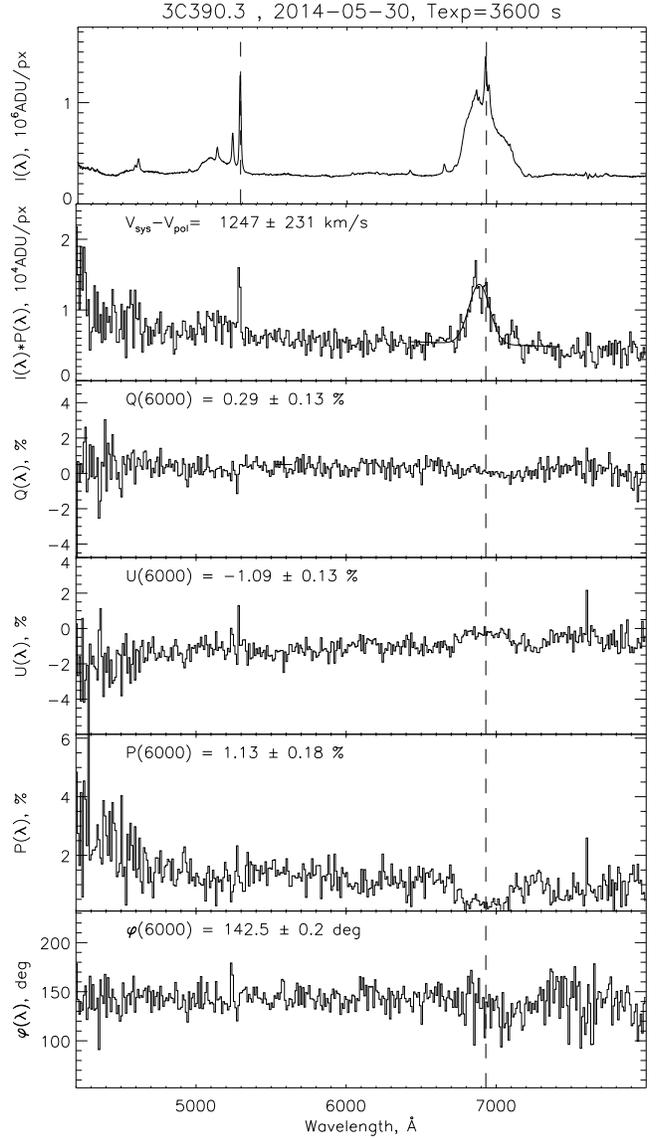}
\caption{The observed spectrum of the unpolarized and polarized flux and its polarization parameters of 3C390.3
on May 30, 2014. From  top to  bottom:
the observed spectrum $I(\lambda)$, the polarized spectrum $I(\lambda)\cdot P(\lambda)$, normalized Stokes
parameters $Q(\lambda)$ and $U(\lambda)$, degree of  linear polarization $P(\lambda)$ and angle of
the polarization plane $\varphi(\lambda)$. Spectra have been corrected for the spectral sensitivity and
ISM polarization. The atmospheric absorption is removed.
} \label{fig1}
\end{figure}

{ We measured  the continuum flux
 at rest wavelength 5100 \AA\, taking an averaged continuum in the H$\beta$ wavelength range.
 To measure  an averaged flux of the polarized continuum, we used a window of the half-width of 500 \AA.
 To avoid the contribution of the [OIII]4959,5007 \AA\, lines we centered the window at 5500 \AA.

 Fig.1 shows
 a typical observed spectrum of the unpolarized and polarized flux and
its polarization parameters of 3C390.3 obtained   on May 30, 2014 in the instrumental
unit without the flux calibration. As it can be seen in Fig. \ref{fig1} there  practically is no difference
between the polarized continuum measured at rest 5100 \AA\ and 5500 \AA\, i.e. the difference is within
 the error-bars.}

From the observed H$\alpha$ and H$\beta$ lines we removed (using Gauss decomposition)
the narrow forbidden lines and obtained the broad components. Then we subtracted
the continuum and obtained the broad emission line fluxes.
For the absolute flux calibration we used the fluxes of
the [OIII]4959+5007 \AA\, ~- forbidden lines, taking the fluxes from \cite{vz91}
(Flux([OIII]=$1.7\cdot 10^{-13}\rm erg/cm^2/s$).
\cite{se02}  found that for mean seeing of $2.5''$, the stellar contribution to continuum in the H$\alpha$ region
 for aperture $2\times11''$ is  $7.0\cdot 10^{-16}\rm erg/cm^2/s/\AA$.
We estimated that the mean continuum flux at the rest wavelength of 5100 \AA\, for the similar
aperture is F(cnt)$ \sim (3.02 \pm 0.63)\cdot 10^{-15}\rm erg/cm^2/s/\AA$.
Thus the contribution of the stellar radiation (host galaxy) to the continuum of
the AGN  3C390.3 is $\sim$20\%. We estimated that this contribution of
the host galaxy is depolarizing the AGN continuum radiation  less than  $<0.25\%$.

\begin{table*}
\begin{center}
\caption[]{Observed Stokes parameters $Q$ and $U$ of the continuum, degree of polarization and polarization angle,
fluxes in continuum, broad lines H$\beta$ and H$\alpha$, polarized broad H$\alpha$ and its shift. The polarized spectra
have been corrected on the ISM polarization.
}\label{tab2}
\begin{tabular}{cccccccccc}

\hline
JD&Q(5500),&U(5500),&$P(5500),$&$\varphi(5500),$&Flux&Flux&Flux&Flux&V$_{pol}$-V$_{sys},$\\
240000+&\%&\%&\%&degrees&Continuum 5100 \AA&Broad H$\beta$&Broad H$\alpha$&Polarized H$\alpha$&km s$^{-1}$\\
\hline
55098&-0.04$\pm$0.24&-1.19$\pm$0.28&1.19$\pm$0.24&134.0$\pm$1.7&4.49$\pm$0.03&3.74$\pm$0.24&2.12$\pm$0.26&2.37$\pm$0.16&~~-800$\pm$300 \\
55394&-0.27$\pm$0.32&-1.35$\pm$0.40&1.38$\pm$0.06&129.4$\pm$5.0&1.97$\pm$0.05&3.11$\pm$0.20&1.26$\pm$0.18&1.08$\pm$0.32&-1559$\pm$441\\
55501&~1.30$\pm$0.28&-0.10$\pm$0.24&1.30$\pm$0.35&177.9$\pm$9.0&2.46$\pm$0.12&2.93$\pm$0.43&1.54$\pm$0.26&2.11$\pm$0.87&-1250$\pm$765\\
55682&~0.85$\pm$0.11&-0.15$\pm$0.10&0.86$\pm$0.11&175.0$\pm$2.0&4.13$\pm$0.05&3.61$\pm$0.18&1.41$\pm$0.18&1.21$\pm$0.27&-1416$\pm$452\\
55713&~1.06$\pm$0.14& 0.50$\pm$0.14&1.18$\pm$0.14&192.7$\pm$2.3&4.19$\pm$0.07&3.86$\pm$0.34&2.27$\pm$0.27&4.02$\pm$0.40&-1294$\pm$402\\
55799&~0.33$\pm$0.17&-1.63$\pm$0.18&1.66$\pm$0.05&140.8$\pm$0.6&2.76$\pm$0.06&4.06$\pm$0.23&2.82$\pm$0.30&7.19$\pm$0.75&-1017$\pm$182\\
55831&~0.52$\pm$0.16&-1.50$\pm$0.14&1.58$\pm$0.19&144.5$\pm$1.7&2.43$\pm$0.07&3.90$\pm$0.24&3.21$\pm$0.55&3.40$\pm$0.39&-1053$\pm$238\\
55885&-0.12$\pm$0.23&-1.57$\pm$0.28&1.57$\pm$0.27&132.9$\pm$9.3&2.68$\pm$0.12&3.56$\pm$0.46&2.84$\pm$0.30&2.92$\pm$1.10&-1273$\pm$630\\
55959&-0.72$\pm$0.08&-1.53$\pm$0.51&1.69$\pm$0.49&122.5$\pm$5.4&3.33$\pm$0.20&4.03$\pm$0.47&1.98$\pm$0.29&1.49$\pm$1.33&-1221$\pm$362\\
55971&~1.13$\pm$0.30&-1.02$\pm$0.29&1.52$\pm$0.19&158.9$\pm$3.5&3.25$\pm$0.12&3.40$\pm$0.41&2.40$\pm$0.28&2.05$\pm$0.92&-1144$\pm$473\\
56032&-0.15$\pm$0.20&-1.49$\pm$0.17&1.50$\pm$0.18&132.0$\pm$2.3&2.76$\pm$0.05&3.62$\pm$0.23&1.44$\pm$0.18&1.44$\pm$0.26&~~-993$\pm$265\\
56064&~0.74$\pm$0.19&-0.83$\pm$0.13&1.11$\pm$0.15&155.8$\pm$2.2&3.05$\pm$0.06&3.25$\pm$0.29&1.15$\pm$0.70&1.20$\pm$0.13&-1323$\pm$237\\
56099&-0.08$\pm$0.12&-0.94$\pm$0.13&0.94$\pm$0.12&132.6$\pm$1.9&3.61$\pm$0.05&3.76$\pm$0.29&1.80$\pm$0.27&1.49$\pm$0.17&-1014$\pm$227\\
56163&~0.10$\pm$0.09&-1.20$\pm$0.20&1.20$\pm$0.19&137.4$\pm$1.9&3.46$\pm$0.04&3.86$\pm$0.31&2.17$\pm$0.31&1.65$\pm$0.35&-1561$\pm$342\\
56181&~0.42$\pm$0.15&-1.02$\pm$0.13&1.10$\pm$0.16&146.3$\pm$2.0&3.62$\pm$0.05&3.67$\pm$0.26&1.87$\pm$0.26&1.46$\pm$0.26&-1585$\pm$272\\
56207&~0.90$\pm$0.16&-1.12$\pm$0.10&1.44$\pm$0.14&154.4$\pm$0.9&3.41$\pm$0.09&4.15$\pm$0.31&1.36$\pm$0.18&1.75$\pm$0.25&-1540$\pm$231\\
56244&~1.02$\pm$0.18&-1.33$\pm$0.15&1.68$\pm$0.18&153.8$\pm$2.0&2.80$\pm$0.05&3.83$\pm$0.28&1.60$\pm$0.22&1.96$\pm$0.30&-1238$\pm$254\\
56329&~0.15$\pm$0.20&-1.17$\pm$0.20&1.18$\pm$0.27&138.7$\pm$2.6&2.94$\pm$0.10&3.57$\pm$0.34&1.30$\pm$0.21&1.27$\pm$0.23&-1644$\pm$284\\
56458&~1.13$\pm$0.15&~0.01$\pm$0.20&1.13$\pm$0.17&180.4$\pm$2.5&2.53$\pm$0.06&3.42$\pm$0.24&1.08$\pm$0.11&0.97$\pm$0.18&-1148$\pm$287\\
56599&~0.21$\pm$0.14&-1.17$\pm$0.14&1.19$\pm$0.06&140.0$\pm$2.1&2.55$\pm$0.06&3.42$\pm$0.30&1.03$\pm$0.13&1.33$\pm$0.19&-1453$\pm$183\\
56713&~0.54$\pm$0.04&-1.39$\pm$0.04&1.49$\pm$0.10&145.6$\pm$1.1&2.53$\pm$0.12&3.30$\pm$0.44&1.46$\pm$0.21&1.08$\pm$0.35&-1474$\pm$431\\
56722&~0.13$\pm$0.05&-1.24$\pm$0.17&1.25$\pm$0.19&138.1$\pm$2.8&2.45$\pm$0.06&3.29$\pm$0.30&1.42$\pm$0.18&1.61$\pm$0.27&-1312$\pm$207\\
56741&~0.13$\pm$0.05&-0.97$\pm$0.11&0.98$\pm$0.14&138.8$\pm$0.5&2.51$\pm$0.06&3.29$\pm$0.32&1.44$\pm$0.19&1.02$\pm$0.26&~~-999$\pm$372\\
56807&~0.29$\pm$0.13&-1.09$\pm$0.13&1.13$\pm$0.18&142.5$\pm$0.2&2.83$\pm$0.05&3.45$\pm$0.27&1.48$\pm$0.16&1.03$\pm$0.21&-1247$\pm$231\\

\hline
\end{tabular}
\\
{UNITS: Continuum 5100 \AA~ in	$10^{-15}\rm erg/cm^2/s/\AA$; Broad H$\beta$ in
$10^{-13}\rm erg/cm^2/s$; Broad H$\alpha$ in $10^{-12}\rm erg/cm^2/s$; Polarized H$\alpha$ in $10^{-14}\rm erg/cm^2/s$.
}
\end{center}
\end{table*}

\section{Results}

Observed polarization parameters in the continuum and broad H$\alpha$ line are given in Table \ref{tab2}, where
we give: the
unpolarized (at 5100\AA\,) and polarized continuum (at 5500 \AA\,) flux, the polarization parameters corrected for the ISM polarization,
the broad H$\beta$ and H$\alpha$ line fluxes, the polarized H$\alpha$ flux
and the difference between the systematic and polarized line velocity
 ($V_{sys}-V_{pol}$) for the broad H$\alpha$ line. Note here that, due to large error-bars in the H$\beta$ wavelength range,
 we did not measure the polarization parameters for
  H$\beta$.

In Fig. \ref{fig1} we show  spectrum observed on May 30, 2014. The slight changes can be noticed in
the rest 23 epochs, but some characteristic structures are the same in all spectra.
 From Fig. \ref{fig1} and Table \ref{tab2}, one can see  several characteristics in unpolarized and polarized spectra of 3C390.3:

\begin{itemize}
\renewcommand{\labelitemi}{$-$}
\item in the unpolarized 3C390.3 spectrum the double-peaked broad emission lines of H$\alpha$ and
H$\beta$ are present; the blue peak was always brighter than the red one during the monitored period,
while the polarized component is single-peaked;
\item the degree of the linear polarization in the continuum was always somewhat higher $\sim2\%$ in the shorter
wavelengths (4000 \AA), changing to $\sim1.5\%$ at 8000 \AA, i.e. there is a trend to increase the linear polarization with decreasing wavelength (i.e. $P\sim{1/\lambda}$);
\item in the polarized light we  observed the H$\alpha$ (sometimes H$\beta$) broad emission line, shifted  to the blue at
$\Delta V=V_{pol}-V_{sys}=-1247\pm231$ km s$^{-1}$ with respect to the systematic velocity;
\item in the H$\alpha$ line the linear polarization $P$ is always less than $\sim$0.2\%
 and has a box-like shape, i.e. P(H$\alpha$) has no pronounced structural features (there are no double peaks).
\end{itemize}

\begin{figure}
\centering
\includegraphics[width=8.3cm]{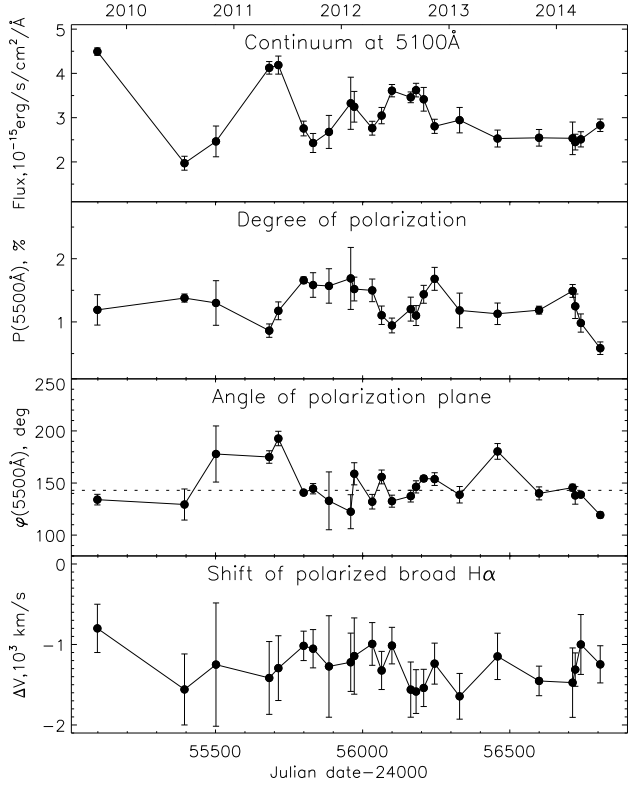}
\caption{From top to bottom: the light curves of the total (unpolarized) continuum flux at 5100 \AA\, ~
in the rest frame, variation of the  degree of the linear polarization $P$ and angle polarization plane  of
the continuum at 5500 \AA\, in the rest frame,  and shift $\Delta$V of the polarized broad H$\alpha$
line relatively to the systematic velocity.} \label{fig2}
\end{figure}

\begin{figure}
\centering
\includegraphics[width=8.3cm]{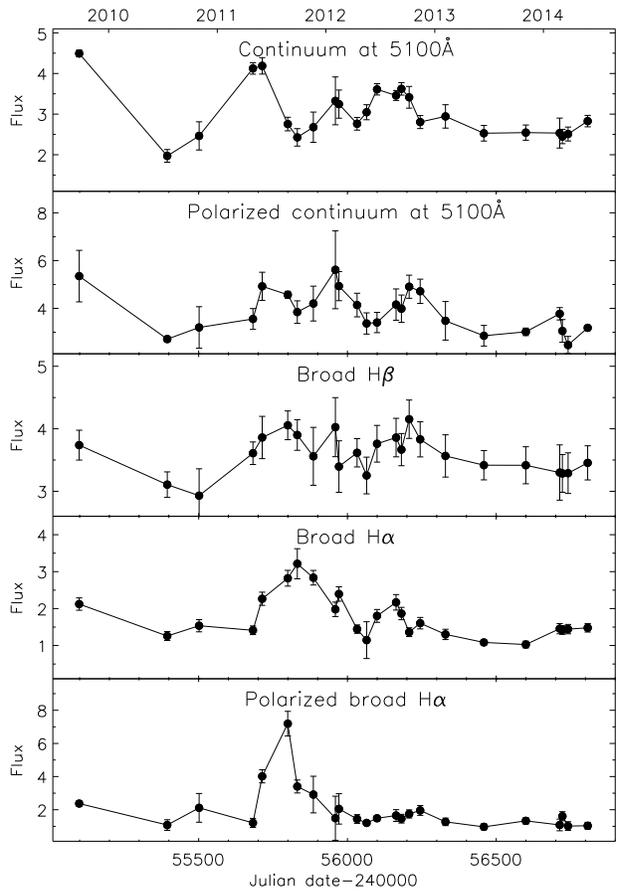}
\caption{3C 390.3 light curves - from  top to  bottom:
 the total and polarized continuum flux at 5100 \AA\, ~
in the rest frame;  the broad H$\alpha$ and H$\beta$ and  the polarized broad H$\alpha$
line. Flux units:
continuum at 5100 \AA\,  in	$10^{-15}\rm erg/cm^2/s/\AA\,$; broad H$\beta$ in
$10^{-13}\rm erg/cm^2/s$; broad H$\alpha$ in $10^{-12}\rm erg/cm^2/s$; polarized H$\alpha$ in $10^{-14}\rm erg/cm^2/s$.
} \label{fig3}
\end{figure}

\subsection{Variability in the continuum and broad H$\alpha$ polarization}

{  The polarization in the continuum has a slight dependence on wavelength
and varied between 1\% and 2\% (see Fig. \ref{fig2}).}
In Fig. \ref{fig2} we present the variability in the continuum at 5100 \AA\, and
polarized continuum measured at 5500 \AA. 
As it can be seen  in Fig. \ref{fig2} and Table \ref{tab2}, the variations in the linear
polarization P\%  sometimes anti-correlate  with the continuum flux,
i.e.  the minimum value of the continuum flux corresponds to the maximum value
of the linear polarization (for example, in the period  26.08.2011-13.11.2012
or JD2455799-JD2456244). The angle of the  polarization plane of the continuum reaches
maximum values ($\sim$180 degrees) in the  period 01. 11. 2010- 01. 06. 2011 or
JD2455501-JD2455713.
{The shift  of the polarized broad H$\alpha$ line had no
significant change during the monitored period (almost within the error-bars).

In Fig. \ref{fig3} the light curves for the total and polarized continuum flux
at 5100 \AA\, are shown\footnote{The polarized continuum flux is obtained taking the continuum flux at 5100\AA\,
and polarization degree at 5500 \AA\,}. A fast inspection of the light curves of
the unpolarized and polarized continuum shows that they are in correlation.

 The variation in the  polarized continuum as well as the variation
 in the broad H$\alpha$, H$\beta$ lines, and polarized
H$\alpha$ line  follows the variation in the unpolarized one, but with some lags between them (Fig. \ref{fig3}).

\begin{table}
\begin{center}
\caption[]{Variability properties of the continuum end emission lines in spectra 3C390.3. UNITS: F(5100)
in ${\rm 10^{-15} erg\ cm^{-2}\ s^{-1}\ \AA^{-1}}$,  F(polarized H$\alpha$) in ${\rm 10^{-14} erg\ cm^{-2}\ s^{-1}\ \AA^{-1}}$,
F(broad H$\beta$) and F(broad H$\alpha$) in ${\rm 10^{-13} erg\ cm^{-2}\ s^{-1}\ \AA^{-1}}$.
}\label{tab3}
\begin{tabular}{lccccc}
\hline \hline
\\
Feature & $<F>$ & $\sigma (F)$ & $\frac{R_(max)}{R(min)}$ & $F_{var}$& $\frac{F(H\alpha)}{F(H\beta)}$ \\
\\
\hline

P(5500), \% & ~~~~2.01 & ~~0.28 & 1.58 & 0.14 & \\
$\varphi$ (5500), deg & 151.80 & 10.90 & 1.33 & 0.07 & \\
F(5100) & ~~~~3.02 & ~~0.63 & 2.28 & 0.21 & \\
F(broad H$\beta$) & ~~~~3.72 & ~~0.72 & 2.36 & 0.19 & \\
F(broad H$\alpha$) & ~~17.72 & ~~3.87 & 3.12 & 0.22 &  4.7\\
F(polarized H$\alpha$)& ~~~~1.96 & ~~1.36 & 7.41 & 0.69 & \\
F(5100) 1995-2007& ~~~~2.31 & ~~1.07 & 5.20 & 0.46 & \\
F(broad H$\beta$) 1995-2007& ~~~~2.39 & ~~0.91 & 4.70 & 0.38 & \\
F(broad H$\alpha$) 1995-2007& ~~~~9.42 & ~~3.36 & 3.40 & 0.35 &3.94 \\

\hline
\end{tabular}

\end{center}
\end{table}

 In Table \ref{tab3} we defined several parameters characterizing the variability of
 the polarized parameters (P\%, $\varphi$), the continuum at 5100 \AA\ and
 the broad H$\beta$,  H$\alpha$  emission line fluxes and the H$\alpha$ polarized flux, using the equation given by \cite{br98}:
         $$F_{var}=\sqrt{\sigma(F)^2-e^2}/F_{mean}$$
where $e^2$ is the mean square value of the individual measurement uncertainty for N observations, i.e. $e^2=\sum e(i)^2/N$.
There, N is the number of spectra, F denotes the mean flux over the
whole observing period, $\sigma(F)$ is the standard deviation, and R(max)/R(min) is the ratio of
the maximal to the minimal  value of the measured  parameters (or flux) in the monitored period.
The parameter $F_{var}$ is an inferred
(uncertainty-corrected) estimation of the variation amplitude with respect to the mean flux.
In Table \ref{tab3} (last 3 lines)  the  variability parameters for the continuum flux
at 5100 \AA\,  and the  broad H$\alpha$ and  H$\beta$ emission line  flux
are given from  the  paper of \cite{sh10}.

As it can be seen from Table \ref{tab3},  during the  monitored period
(2009-2014) the average fluxes and their amplitudes in broad lines
and continuum were about  2 times  bigger  while $F_{var}$ (variability)
was 2 times smaller than in the period of spectral monitoring (1995-2007).
The variability $F_{var}$ for polarization parameters is not high
(14\% for  P, 7\% for $\varphi$).  However,  for the polarized  H$\alpha$  line $F_{var}$
is $\sim69\%$  that is considerably larger than the variability of
another parameters. {The polarization in H$\alpha$ is close to the zero level, that could be reason for the
large variability.}
 The maximum  of the polarized H$\alpha$ flux
 ($F=(7.19\pm0.75)\cdot 10^{-14} {\rm erg\ cm^{-2}\ s^{-1}}$ (Table \ref{tab2}) was observed on
 26. 08. 2011. At this time
 the broad H$\alpha$ and H$\beta$ emission line and polarized continuum
at 5100\AA\,  were close to have the maximum  fluxes, but the unpolarized (total) continuum flux was maximal
in 01. 05.-01. 06. 2011, i.e.,  60-90 days before the maximum in the line flux. (Table \ref{tab2} and Fig. \ref{fig3}).
Obviously, we see a lag of response of the broad line to the continuum flux (see Fig. \ref{fig3}).

\begin{figure*}
\centering
\includegraphics[width=16cm]{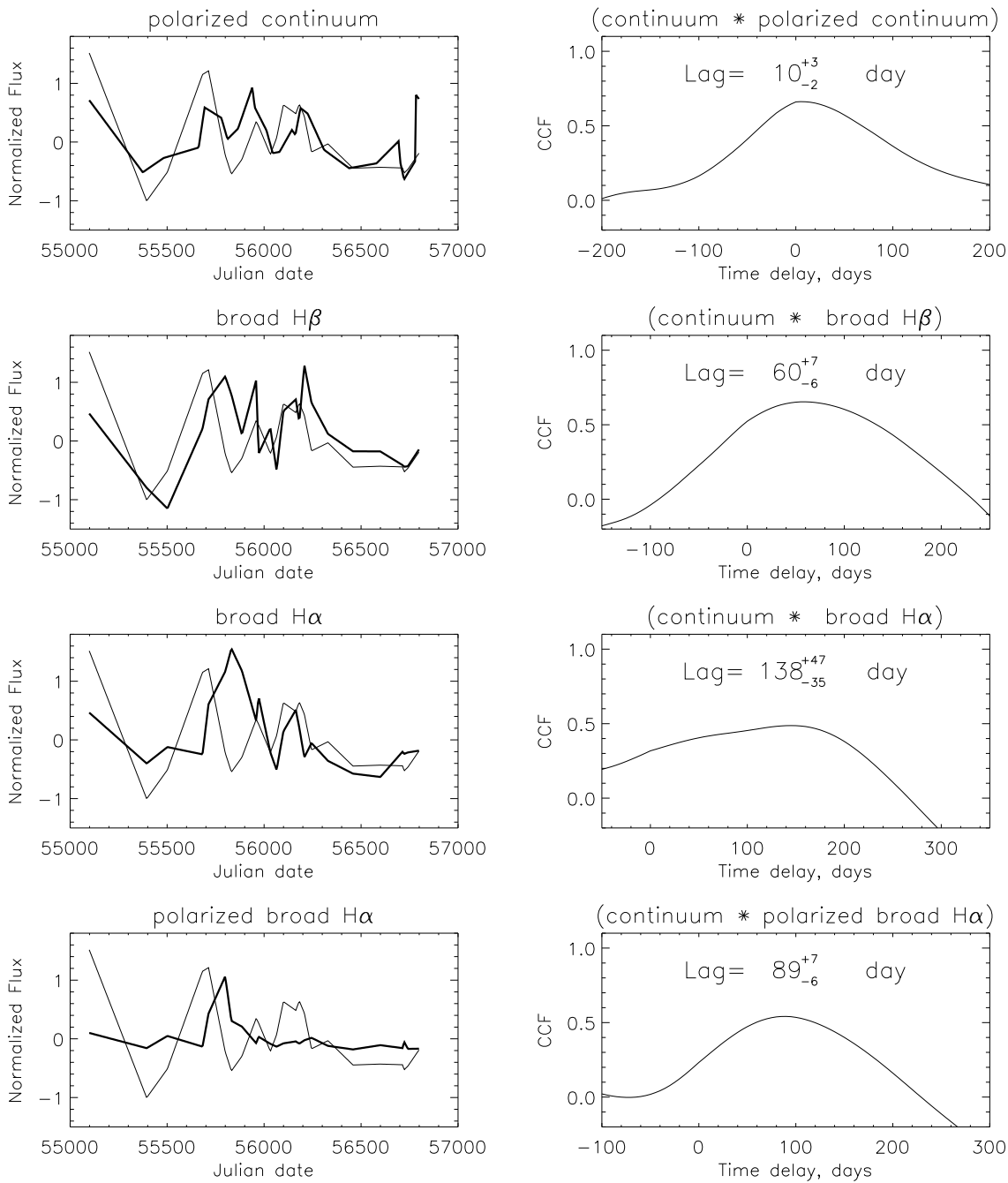}
\caption{Left -- the interpolated normalized light curves, from top to bottom  (thick lines): the polarized continuum,
the broad H$\beta$ flux,
the broad H$\alpha$ flux,  the polarized H$\alpha$ flux compared with  the unpolarized
continuum (thin lines). Right --  from top to
bottom the lags between the unpolarized and polarized continuum at rest 5100 \AA;   the broad H$\beta$ line, the broad H$\alpha$
line, and the  polarized H$\alpha$ fluxes.
} \label{fig4}
\end{figure*}

\subsection{CCF analysis - dimension of the scattering regions}

Using the reverberation method, which is the search for correlations between the broad emission line and continuum
flux variations, it is possible to study geometry and dynamics of the BLR \citep[see][and references therein]{pp93}.
 By analogy  we used a long-term
 spectropolarimetric monitoring of 3C390.3 to study the properties of the scattering gas geometry by reverberation method.
To do this we  applied the CCF analysis using first the interpolation method  \citep[so called ICCF, see][]{gs86}. We interpolated the light curves in the
unpolarized and polarized continuum (as well as in the H$\alpha$ and H$\beta$ lines), after that we used the CCF
and computed the lags relative to the CCF peak.

In Fig. \ref{fig4} (left) we plot   the interpolated light curves of (from  top to  bottom)
 the polarized continuum, the broad  H$\beta$,  the broad  H$\alpha$   and  the polarized
  H$\alpha$  (thick lines) compared with the interpolated normalized light curves for the unpolarized continuum (thin line).
  As it can be seen in  Fig. \ref{fig4} (left)  different lags
between the unpolarized continuum (thin line) and mentioned fluxes are present.

Since different  time series analyses can give different lags for the same set of data
\citep[see in more details][]{ko14}, therefore we calculated the lags using Z-transformed Discrete Correlation Function
\citep[ZDCF, see][]{al13}
and SPE - Stochastic Process Estimation \citep{zu11} in addition to the ICCF method. The estimated lags are given in Table \ref{tab4}.

We found  that  the lags between unpolarized and polarized continuum at 5100 \AA\,~ are $\sim$10 -- 40 days;
  between the unpolarized continuum and the broad H$\beta$ line $\sim$ 60 -- 80 days; between the unpolarized continuum
   and broad H$\alpha$ flux $\sim$140 -- 190 days; between the unpolarized continuum and polarized H$\alpha$ is
   $\sim$90 -- 160 days (see Table \ref{tab4}).
In all three methods we obtained that the lag
between the unpolarized and polarized continuum is significantly smaller than   lags between the continuum and broad lines,
that indicates that the {  variable} polarized continuum  {  is originating}
in a region which is significantly smaller than the BLR.

\begin{table}
\begin{center}
\caption[]{The lags obtained using different methods: ICCF -- interpolation method \citep{gs86},
ZDCF -- Z-transformed Discrete Correlation Function \citep{al13}
and SPE - Stochastic Process Estimation \citep{zu11}.}\label{tab4}
\begin{tabular}{lccc}
 \hline
lag between: &                                      ICCF   &                ZDCF &  SPE   \\
\hline
&&&\\
cnt - pol-cnt         &               10$_{-2}^{+3}$  &    27$_{-10}^{+9}$ & 39$_{-9}^{+7}$   \\
&&&\\
cnt - broad H$\beta$   &              60$_{-8}^{+7}$  &        79$_{-15}^{+16}$ & 79$_{-9}^{+8}$    \\
&&&\\
cnt - pol. broad H$\alpha$  &         89$_{-9}^{+7}$  &     128$_{-29}^{+32}$ & 156$_{-21}^{+8}$\\
&&&\\
cnt - broad H$\alpha$       &        138$_{-35}^{+47}$ &  184$_{-12}^{+28}$ & 186$_{-6}^{+7}$  \\
\ &&&\\
\hline
\end{tabular}

\end{center}
\end{table}

Note here that our lag  estimates are in  good agreements with these given in literature, as e.g. \cite{sh10} obtained lags
 $127^{+18}_{-18}$ days  for H$\alpha$   and   $93^{+20}_{-18}$ days for H$\beta$. 

\begin{figure}
\centering
\includegraphics[width=8.3cm]{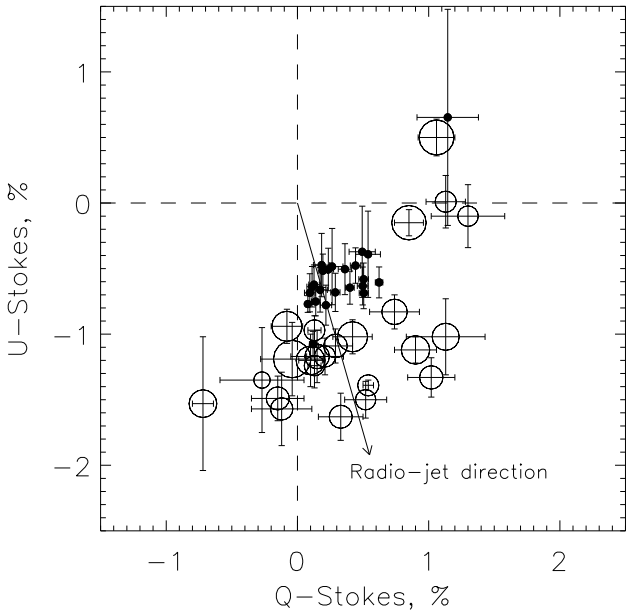}
\caption{The variation of the continuum (open circles) in the Stokes parameters (UQ) space
with the projection of the radio jet direction (arrow).
The dimension of  circles corresponds to the intensity continuum flux at 5100 \AA\,  in the rest frame.
The arrow represents the direction of the radio jet of 3C390.3 in the UQ space.
The black full circles represent the H$\alpha$ broad line Stokes parameters measured in the frame of the
broad line profile.} \label{fig5}
\end{figure}

\section{Discussion}

\subsection{Polarized continuum }

 Using the results of our spectropolarimetric monitoring of 3C 390.3  (data for 24 epochs during 5 years) we
 found the lag $\sim10-40$ days between the unpolarized and polarized continuum flux variations. The estimated size
 of the BLR is significantly larger (lag $\sim 60-190$ days). This indicates that the region {  of polarized radiation} which
 contributes to the variability of the polarized continuum is
more compact  than the BLR in the case of 3C390.3. A similar result  (lag $\sim2$ days) is obtained for Mrk 6 \citep[][]{af14},
where the BLR is estimated to be significantly larger than the scattering region. \cite{ga12} found that  in the case of
NGC 4151 the lag between the
unpolarized and polarized continuum flux variations is $\sim8$ days and almost the same as the BLR lag, however
it seems that their error-bars
of the estimated lags are large \citep[see their Table 1 in][]{ga12}

The average  polarization angle ($\varphi=152 \pm 11$ degrees) of  the continuum at rest wavelength 5500 \AA\,
is  consistent  and almost aligned with the radio jet
 angle $\sim144^\circ$ \citep[][]{lp91}.  The variability of  $\varphi$  is relatively small ($F_{var}\sim7\%$, Table \ref{tab3}),
 but the degree of the linear polarization of the continuum changes as $F_{var}\sim14\%$, that may indicate  an additional
 continuum component to that in the nucleus.
This is in favor of the results obtained by \cite{ar10}. According to the results of VLBI monitoring
 there are the observational
 evidence for the connection between the variable optical continuum nucleus of 3C390.3 and the compact
 radio emission of the jet on the sub-parsec scale \citep[][]{ar10}. To explain this correlation, \cite{ar10} suggested that
 the variable optical continuum emission is generated in the innermost part of the jet.

{  The polarization in the 3C 390.3 continuum  probably has three sources:
(i) We expect to have
the polarization from the disc (light scattering in a plane-parallel disc atmosphere), with perpendicular 
polarization angle to the jet direction (and almost constant);
(ii) Scattering on the inner part of the torus, where the vector of polarization is aligned
with the radio-jet (without fast variability in the continuum polarization); and (iii)
The synchrotron continuum emission of the jet 
 (whose polarization vector is approximately 
 perpendicular to the jet direction) that probably contributes to the variability in the continuum polarized light.}
 This also can 
 be seen in Fig. \ref{fig5} where the variability in the continuum (open circles) has  arc-like structure with 
 respect to the jet direction (denoted  with an arrow). The jet angle is taken from \cite{al88,al96}.
 
{  The observed decrease of the polarization in the continuum as a function of wavelength
  may be caused by the Rayleigh-scattering on the torus or the inverse Compton 
effect \citep[][]{bs87} in the relativistic plasma 
jet. However, a detailed study of this effect requires modeling and it is out of the scope of this article.}
Another possible explanation for decreasing polarization may be wavelength
dependent dilution by the host galaxy continuum, which will be redder than the
AGN. However as we noted above the contribution of the host galaxy to the unpolarized
flux is $\sim$ 20\%, and  the estimated contribution to the
polarization is around 0.25\%. This cannot cause the observed changes in the optical
polarization, which are about  1\%.

\begin{figure}
\centering
\includegraphics[width=8.3cm]{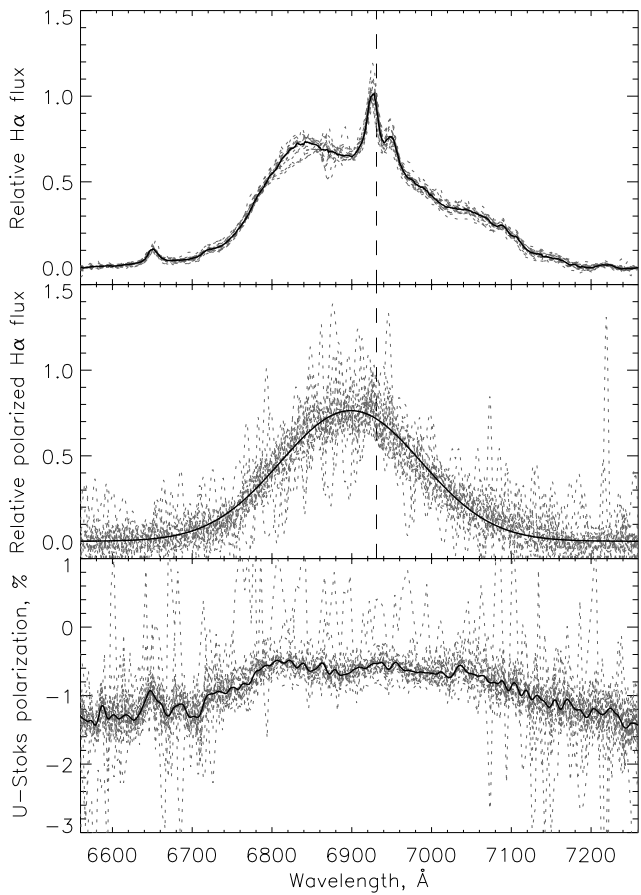}
\caption{ From  top to  bottom: the broad H$\alpha$ line, polarized  H$\alpha$ and polarization parameter $U$.
Faint dashed lines represent observations from different epochs and solid bold lines represent averaged profiles
of the unpolarized and polarized
H$\alpha$ and parameter $U$} \label{fig6}
\end{figure}

\subsection{Polarized  H$\alpha$  }

There are several models proposed to explain the two-peak structure of the broad Balmer emission   lines in 3C 390.3. 
A widely accepted model is the formation of the broad lines in a relativistic disc. However, there is some contradictions 
between these models and the polarization observations of 3C 390.3. 
\cite{sm04,sm05}   proposed {  two mechanisms of polarization in the optical
  continuum and H$\alpha$ broad emission lines in Sy1 nuclei -- 
equatorial (scattering on the inner part of the torus) and polar (scattering on the jet) polarization mechanisms.}
 Both scattering components should be present in all 
  broad line AGNs, and  their polarization properties can be broadly understood in terms of an orientation effect 
  \citep[Unified model, see]
  []{an96}.
{  In the case of equatorial scattering model, the broad emission lines which are emitted from a rotating  disc (or outer regions of 
accretion disc-BLR)  have to show a double-peaked structure in the polarized light, similar as in the unpolarized light. 
 The degree of polarization will 
 be maximal in the broad line wings and  a minimum is expected in the line center.}
 The  angle $\varphi$ of plane  polarization is aligned with the projected disc rotation axis and hence with the 
 radio source axis.
 
In the case of 3C 390.3, we observed  an unexpected difference between the unpolarized and polarized broad 
line profile of  H$\alpha$: the unpolarized H$\alpha$  line profile has
 double-peaked structure with the blue peak at  $V_r\sim-3500$ km s$^{-1}$ and red one at $V_r\sim+5000$ km s$^{-1}$. 
 But the polarized H$\alpha$ line profile is single-peaked and shifted  to blue at  $V_r\sim-1200$ km s$^{-1}$  with respect to the
 narrow component (or systematic velocity). Also,  the degree of the linear polarization 
  of H$\alpha$ is small (around 0.1\% -- 0.2\%, close to zero) and has a
  box-like shape without any significant
    structural details, i.e. it seems that the broad H$\alpha$ line is almost completely depolarized.
    This is contrary to the equatorial
  scatter model. On the other hand in 3C 390.3  polarization angle ($\varphi$) is parallel to the
  pc-jet axis as it is expected in the equatorial scatter model (see Fig. \ref{fig5} - full circles).

\begin{figure}
\centering
\includegraphics[width=8.3cm]{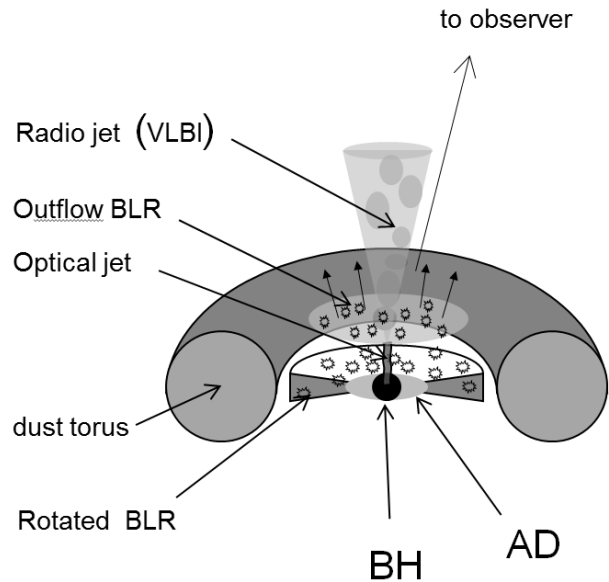}
\caption{Sketch illustrating a possible central region of 3C 390.3. {  Two-component BLR:
the double-peaked lines originate in the disc-like region, and depolarization is by warm clouds in an outflowing part of the BLR.}
The part of the variable polarized continuum seems to be emitted from the jet.  } \label{fig7}
\end{figure}

It should be noted that   \cite{co00} considered a model of 3C 390.3 in which H$\alpha$  photons emitted by a biconical 
flow are scattered by the inner wall of the torus. Thus, the scattering plane is perpendicular to the radio jet (e.g. 
E-vector parallel to pc-jet axis) and produces a single-peaked scattered H$\alpha$ line profile. However, 
this model does not agree with the optical monitoring data for the BLR, where  the double-peaked broad emission line is formed.  
The CCF-analysis for the H$\beta$ and H$\alpha$ line wings in 3C 390 shows that there is no
 a significant delay in the variation of the line wings with respect to the central part of the line, or relative to each other.
  The flux in the H$\beta$ line wings and core also varied simultaneously. This result indicates a 
  dominant circular motion in the BLR and it is in favor of a model in which the main contribution to the broad line fluxes 
  originates in the accretion disc and not from the biconical flow \citep[][]{di98,di12,sh10,po11,zh11,zh13}.

\subsubsection{Polarized profile of the broad H$\alpha$ line}

 In Fig. \ref{fig6} we present the main  unpolarized (first panel) and polarized (second panel)  broad H$\alpha$ shapes. 
 The polarized and unpolarized lines are normalized on their intensities
  in order to compare the profile variability. 
As it can be seen in Fig. \ref{fig6} there is no  big change in the polarized and unpolarized line profile. The unpolarized H$\alpha$
line has a double-peaked profile, 
the peaks are located (within the error-bars)  as earlier measured \citep[see e.g.][]{po11}: 
the blue peak is located at  
V$_b \sim -3500 $ km s$^{-1}$ while 
the red peak had V$_r\sim 5000$ km s$^{-1}$ for H$\alpha$. Contrary, the polarized H$\alpha$ is single-peaked and 
shifted to the blue for about -1200 km s$^{-1}$. 

Also, in the third
panel of Fig. \ref{fig6} we present the Stokes $U$ parameter. It is interesting that the  $U$ parameter has a box-like shape,
that indicates presence
of a depolarization region, i.e. presence of the warm gas between an observer and accretion disc.

The Stokes parameters for the broad line are present in Fig. \ref{fig5} as black full circles. 
It is interesting that, they are following the jet direction. It means that
there is a slight polarization from the disc, but in  contrast to Mrk 6 \citep[see][]{af14},
 the effect of the equatorial scattering  cannot be seen.

\subsection{Polarization - Connection between disc and jet}

To speculate about the possible model that can explain the unpolarized and polarized H$\alpha$ line profile and the continuum, 
let us recall several observational facts:

(i) 3C390.3 is a powerful radio source with two lobes, FRII radio galaxy   with relatively strong compact core.
There are two extended lobes in position angle $PA\sim144^{\circ}$, separated by $\sim223''$ each with a hot spot at the
end \citep[][]{lp91}. Also, there is a weak well collimated thin jet in $PA\approx -37^{\circ}$, linking the core with the northern
lobe \citep[][]{lp95}. In this parsec-jet the VLBI observations at 5 GHz detected superluminal motion (with $v/c\sim4$)
\citep[][]{al88,al96}. 

(ii) It seems that a part of the optical continuum is coming from the jet since the optical continuum and radio emission from a 
jet-structure are correlated \citep[][]{ar07,ar10}

(iii) The unpolarized H$\alpha$ shape and its variability are in agreement with the disc emission model \citep{sh10,po11}. 
However, there is a central component shifted to the red 
 from 300 km s$^{-1}$ to 800 km s$^{-1}$ and additionally the broad double-peaked H$\alpha$ 
 line shows a blue shift, indicating a wind at the
  disc surface with velocities from -300 km s$^{-1}$ to -800 km s$^{-1}$ \citep[][]{po11}. 
  Contrary, the polarized H$\alpha$ broad line profile 
  is single-peaked with the blue shift of 1200 km s$^{-1}$, that is not in the favor of the disc
 model \citep[][]{sm04,sm05}, but seems to agree with two-side outflow model \citep{co98,co00} 

To explain all these observational facts, one can take into account a complex BLR model (see Fig. \ref{fig7}), where 
the disc-like BLR is covered by {  an outflowing region (the wind with speed around -1200 km s$^{-1}$)
-- with a number of warm clouds which also can emit the H$\alpha$ line. This region can  
depolarize the  disc-like emission. The two-component BLR (disc covered by an outflowing region of 
randomly distributed clouds) seems to be 
a good approximation for nearby AGN \citep[][]{kz13}, as well as for a number 
of single-peaked AGN \citep[see in more details][]{bo09}.}

  One can expect that {  the outflowing region with warm gas} can contribute to the additional 
electron scattering, which may actually cause the
line emission to be polarized. However, in the outflowing region, the polarization vector is parallel to the disc plane,
and practically
depolarizes the polarized light from equatorial scattering. 

{   In a scenario of the two-component model 
\citep[disc+outflowing components][]{po11}, the blue-shifted H$\alpha$ component is
coming from  the outflowing part of the BLR, and cannot be (self)depolarized, while the disc-like component is
depolarized by this region.

Moreover, the two-component BLR model is able to explain both,
 the double-peaked unpolarized emission and  depolarization of the broad H$\alpha$ line 
 (box-like structure in the polarized line profile, see Fig. \ref{fig1}). 
 Alternatively, the hot depolarization gas may have an inflowing component, that is in contradiction
with results obtained in the long term profile line analysis \citep[see][]{po11}.}
 
 The part of the continuum is originating in this jet like part (outflow) and it may be the reason for
 the wavelength dependent polarization in the continuum, and relatively high level of the
continuum polarization \citep[around 2\%, in comparison with Sy1 that is around 1\%, see][]{af14}.

\section{Conclusion}

We presented results of the spectropolarimetric monitoring of 3C390.3, obtained  at 24 epochs in
the period of 2009-2014. The galaxy has been observed
with 6-m telescope of SAO observatory using SCORPIO-2 instrument with
spectral resolution of 7-10 \AA\, in the spectral range between 4000 \AA\, and 8000 \AA\,.
 Also, we observed and estimated  the contribution of the ISM polarization to the observed 3C 390.3
 polarization. We measured polarization parameters for the continuum at 5100\AA\, (rest wavelength) and H$\alpha$
 line  and their variabilities, and explored the lag between the unpolarized and polarized flux.
 On the basis of our investigation we can outline the following conclusions:

i) During the 5-year monitoring period we found variation in the broad line and continuum polarization parameters.
We found a lag of $10-40$  days between the polarized
and unpolarized  continuum flux variation at 5100\AA\,. This lag is significantly smaller than one we
found for the BLR (lag for H$\beta$ is $60-80$  days  and for H$\alpha$ is $140-190$ days) i.e.
the scattering region of the continuum  probably is not the BLR (accretion disc).
It seems that the polarized continuum has a component which is coming from the disc (E vector is orientated in
the jet direction), and another, that contributes to variability
 that may be synchrotron contribution to the continuum from a pc-jet component.

ii) The unpolarized double-peaked broad emission   H$\beta$ and H$\alpha$ lines are observed as single peaked polarized lines.
The polarized line
is shifted to the blue for about -1200 km s$^{-1}$, and such profile was  not expected in the case of the equatorial scattering.
Additionally, we observed a box-like profile
of the line polarization (that has a small value $\sim$0.1\%), that indicates depolarization by the warm gas.
Taking into account the results from
the modeling of variability, we proposed that depolarization
region is {  an outflowing region} located above the disc-like emission region (that
emits double-peaked lines) and plays a role in the disc line depolarization.

\section*{Acknowledgments}
The results of observations were obtained with the 6-m BTA telescope
of the Special Astrophysical Observatory of Academy of Sciences,
operating with the financial support of the Ministry of Education
and Science of Russian Federation (state contracts no.
16.552.11.7028, 16.518.11.7073).The authors also express appreciation
to the Large Telescope Program Committee of the RAS for
the possibility of implementing the program of Spectropolarimetric
observations at the BTA. This work was supported by the Russian
Foundation for Basic Research (project N12-02-00857) and
the Ministry of Education, Science and Technological Development (Republic of Serbia) through
the project Astrophysical Spectroscopy of Extragalactic Objects
(176001).  We thank to M. Gabdeeev for help in spectropolarometric observations.
L.\v C. Popovi\'c thanks to the
COST Action MP1104 'Polarization as a tool to study the Solar
System and beyond' for support. {  We thank to an anonymous referee for very useful comments.}

\end{document}